\documentclass[prd,nofootinbib,preprint,superscriptaddress]{revtex4}

\usepackage{amsmath}
\usepackage[dvips]{graphicx}

\newcommand{\capdef}{}
\newcommand{\mycaption}[2][\capdef]{\renewcommand{\capdef}{#2}%
       \caption[#1]{{\footnotesize #2}}}
\makeatletter
\renewcommand{\fnum@table}{\textbf{\tablename~\thetable}}
\renewcommand{\fnum@figure}{\textbf{\figurename~\thefigure}}
\makeatother

\newcommand {\be}{\begin{equation}}
\newcommand {\ee}{\end{equation}}
\newcommand {\ba}{\begin{eqnarray}}
\newcommand {\ea}{\end{eqnarray}}

\begin{document}


\vspace*{10mm}

\title{Reconciling results of LSND, MiniBooNE and other experiments 
with soft decoherence \vspace*{1.cm} }

\author{\bf Yasaman Farzan}
\email{yasaman_at_theory.ipm.ac.ir} \affiliation{Institute for
research in fundamental sciences (IPM), PO Box 19395-5531, Tehran,
Iran}

\author{\bf Thomas Schwetz}
\email{schwetz_at_cern.ch} \affiliation{Theory Division, Physics Department,
CERN, 1211 Geneva 23, Switzerland}

\author{\bf Alexei Yu Smirnov}
\email{smirnov_at_ictp.it} \affiliation{ International Centre for
Theoretical Physics, Strada Costiera 11,
34014 Trieste, Italy, \\
 Institute for Nuclear Research, Russian Academy of Sciences,
Moscow, Russia \vspace*{1.cm}}

\preprint{CERN-PH-TH/2008-102}

\begin{abstract}
  \vspace*{.5cm} We propose an explanation of the LSND signal via
  quantum-decoherence of the mass states, which leads to damping of
  the interference terms in the oscillation probabilities.  The
  decoherence parameters as well as their energy dependence are chosen
  in such a way that the damping affects only oscillations with the
  large (atmospheric) $\Delta m^2$ and rapidly decreases with the
  neutrino energy.  This allows us to reconcile the positive LSND
  signal with MiniBooNE and other null-result experiments.  The
  standard explanations of solar, atmospheric, KamLAND and MINOS data
  are not affected. No new particles, and in particular, no sterile
  neutrinos are needed. The LSND signal is controlled by the 1-3
  mixing angle $\theta_{13}$ and, depending on the degree of damping,
  yields $0.0014 < \sin^2\theta_{13} < 0.034$ at $3\sigma$. The
  scenario can be tested at upcoming $\theta_{13}$ searches: while the
  comparison of near and far detector measurements at reactors should
  lead to a null-result a positive signal for $\theta_{13}$ is
  expected in long-baseline accelerator experiments. The proposed
  decoherence may partially explain the results of Gallium detector
  calibrations and it can strongly affect supernova neutrino signals.
\end{abstract}

\maketitle

\section{Introduction}

Reconciling the LSND signal for $\bar\nu_\mu\to\bar\nu_e$
transitions~\cite{Aguilar:2001ty} with the by now fully
established existence of neutrino oscillations~\cite{sk-atm,
MINOS, K2K, sno, kamland} on the one hand side and bounds on
oscillations~\cite{Apollonio:2002gd, Boehm:2001ik, karmen,
Dydak:1983zq, Declais:1994su, NOMAD, NuTeV} on the other is a
long-standing problem in neutrino physics. Recently the MiniBooNE
experiment~\cite{MB} failed to confirm the oscillation
interpretation of the LSND signal \cite{0805.1764}, putting
further pressure on its interpretation. Usually the LSND result is
considered as an indication for sterile neutrino oscillations,
despite difficulties of the corresponding models to explain the
global data~\cite{Maltoni:2002xd} including cosmological
observations (see Ref.~\cite{Maltoni:2007zf} for a recent analysis
including the MiniBooNE results).
Triggered by these problems many ideas and scenarios have been proposed in
order to explain LSND, some of them involving very exotic physics.
This  includes sterile neutrino decay~\cite{Ma:1999im,
Palomares-Ruiz:2005vf}, violation of the CPT~\cite{cpt} and/or
Lorentz~\cite{lorentz} symmetries, mass-varying
neutrinos~\cite{MaVaN}, short-cuts of sterile neutrinos in extra
dimensions~\cite{Pas:2005rb}, a non-standard energy dependence of
the sterile neutrino parameters~\cite{Schwetz:2007cd}, or sterile
neutrinos interacting with a new gauge boson~\cite{Nelson:2007yq}.

In the present paper, we revisit the possibility that the origin of
the LSND signal might be quantum decoherence in neutrino
oscillations~\cite{Gabriela-04, Gabriela-06}.  Such effects can be
induced by interactions with a stochastic environment; a possible
source for this kind of effect might be quantum gravity \cite{Hawking,
Ellis:1983jz, Giddings}, see Ref.~\cite{mavromatos} for a recent
discussion. We take a phenomenological approach and determine the form
and magnitude of the new effects from observations without reference
to possible origins of the decoherence.
Neutrino oscillations, being a quantum interference effect over
macroscopic distances, provide a sensitive test for decoherence.  The
possibility to use neutrinos as a probe of quantum decoherence has
been explored for atmospheric neutrinos~\cite{Lisi}, solar
neutrinos~\cite{Fogli}, KamLAND~\cite{Fogli, Schwetz:2003se}, future
long-baseline experiments~\cite{lbl, Sakharov}, and neutrino
telescopes~\cite{high-e}. See also Ref.~\cite{other-decoh}.
Other quantum systems which have been considered to search for
decoherence include, for example, $K\bar{K}$~\cite{Ellis:1983jz,
kaons, kaons-ex} and $B\bar{B}$~\cite{Bertlmann:1997ei,Go:2007ww}
oscillations or neutron interferometry~\cite{Benatti}.  Throughout
this paper we will assume that quantum decoherence affects only the
neutrino sector.

Previous attempts to explain the LSND signal by quantum
decoherence~\cite{Gabriela-04, Gabriela-06} seem to be in conflict
with the present data.  Indeed, for the set-up of NuTeV with a
baseline of $\sim 1$ km and the average energy of 75 GeV the model in
\cite{Gabriela-06} predicts $P(\nu_\mu \rightarrow \nu_e) =
P(\bar{\nu}_\mu \rightarrow \bar{\nu}_e) = 3\times 10^{-3}$ while the
one in \cite{Gabriela-04} yields $ P(\bar{\nu}_\mu \rightarrow
\bar{\nu}_e) = 0.2$.  Both of these predictions strongly violate the
upper bound on the neutrino and anti-neutrino oscillation
probabilities from NuTeV: $P(\nu_\mu \rightarrow \nu_e) ,
P(\bar{\nu}_\mu \rightarrow \bar{\nu}_e) < 5 \times 10^{-4}$
($90\%$~C.L.)~\cite{NuTeV}.  Furthermore, the model of
\cite{Gabriela-04} (where in addition to decoherence, also
CPT-violation is also introduced) cannot account for the spectral
distortion in the anti-neutrino signal observed by KamLAND. The
scenario of \cite{Gabriela-06} is also disfavored by the absence of a
signal in KARMEN~\cite{karmen}, NOMAD~\cite{NOMAD} and
MiniBooNE~\cite{MB}.
In the present paper we propose a different set of decoherence
parameters with a special energy dependence. As a result, the
problems outlined above can be avoided. In our scenario, the
decoherence effects become more significant with decreasing the
energy; thus, we refer to this scenario as ``{\it soft
decoherence}".

The paper is organized as follows. In sec.~\ref{sec:scenario}, we
describe our scenario of the decoherence effect and discuss how
the LSND signal is explained. In sec.~\ref{sec:other}, we show
that by selecting suitable parameters, within this scenario the
LSND result can be reconciled with all other oscillation results
(both positive and negative). In sec.~\ref{sec:future}, we discuss
how this scenario can be checked in forthcoming and future
experiments. Discussions and conclusions will be presented in
sec.~\ref{sec:conclusions}.

\section{The scenario: decoherence and the LSND result}
\label{sec:scenario}

\subsection{Soft decoherence}

Let us describe the neutrino system by the density matrix $\rho$ in
the mass state basis, $\nu_i$, $i = 1,2,3$.  The decoherence effects
in the evolution of the density matrix can be parameterized by
introducing a new term $\mathcal{D}[\rho]$ as
\be\label{modified-evolution}
  \frac{d\rho}{dt}=-i[H,\rho]-\mathcal{D}[\rho] \, .
\ee
This term violates the conservation of Tr($\rho^2$) and, hence,
leads to the evolution of pure states into mixed states. As
mentioned in introduction, such a term can be induced by
interaction with a stochastic environment; a possible source for
this kind of effect might be quantum gravity \cite{Hawking,
Ellis:1983jz, Giddings}. The form of the operator
$\mathcal{D}[\rho]$ can be constrained by imposing some general
requirements on the evolution of the system. First, {\it complete
positivity} implies the so-called Lindblad form for
$\mathcal{D}[\rho]$ \cite{Lindblad, Banks}:
\be {\mathcal{D}}[\rho]=\sum_n \left[ \{ \rho,D_n D_n^\dagger
\}-2D_n\rho D_n^\dagger\right] \,. \ee
where $D_n$ are some general complex matrices.  This form arises
from ``tracing away'' the dynamics of the
environment~\cite{Giddings, Adler}. Second, with general complex
$D_n$ unitarity is violated, {\it i.e.,} $d{\rm Tr}(\rho)/dt$ can
be nonzero. Therefore, we require that $D_n$ are Hermitian,
$D_n^\dagger = D_n$.  In addition to $d{\rm Tr}(\rho)/dt = 0$, the
Hermiticity of $D_n$ guarantees that the entropy [{\it i.e.,}
$S(\rho)= -{\rm Tr}(\rho \ln\rho)$] cannot decrease~\cite{Banks,
Benatti:1987dz}. Finally, we require that the average energy of
the system, Tr($\rho H$), is conserved. It is straightforward to
check that this can be achieved by demanding $[H,D_n]=0$. In other
words, unitarity and conservation of the energy-momentum imply
that $D_n$ and $H$ can be simultaneously diagonalized. In the
neutrino mass basis, we can therefore write
\be
H={\rm Diag}[h_1,h_2,h_3]  \ \ \ \ {\rm and} \ \ \ \ D_n={\rm
Diag}[d_{n,1},d_{n,2},d_{n,3}] \ ,
\label{form}
\ee
where $h_i^2 \equiv p^2+m_i^2$, and $d_{n,i}$ are real quantities of
dimension [mass]$^{1/2}$ whose energy-dependence is unknown.
In this paper  we adopt a phenomenological approach and
determine  $d_{n,i}$ from observations without discussing their
possible origins.

Solving the evolution equation  Eq.~(\ref{modified-evolution}),
with $H$ and $D_n$ given in (\ref{form}) we find
\ba \label{evolved} \rho(t)=
 \left[ \begin{matrix}
 \rho_{11}(0)&\rho_{12}(0) e^{-(\gamma_{12}-i\Delta_{12})t}&
 \rho_{13}(0)e^{-(\gamma_{13}-i\Delta_{13})t}
\cr \rho_{21}(0) e^{-(\gamma_{21}-i\Delta_{21})t}&
\rho_{22}(0)&\rho_{23}(0) e^{-(\gamma_{23}-i\Delta_{23})t}\cr
\rho_{31}(0) e^{-(\gamma_{31}-i\Delta_{31})t}&\rho_{32}(0)
e^{-(\gamma_{32}-i\Delta_{32})t}&\rho_{33}(0)
\end{matrix} \right],
\ea
where $\rho_{ij}(0)$ are the elements of the density matrix at the
initial moment,
\be
\gamma_{ij} \equiv \sum_n(d_{n,i}-d_{n,j})^2 \ \ {\rm and} \ \
\Delta_{ji} \equiv  h_j-h_i \approx \frac{\Delta m_{ji}^2}{2E_\nu} \,.
\ee
Notice that $\gamma_{ij} = \gamma_{ji}$ whereas
$\Delta_{ij} = -\Delta_{ji}$, and the diagonal elements of $\rho$ do not
depend on  time.

Let us consider the transitions between the flavor states,
$\nu_\alpha = \sum_{i} U_{\alpha i} \nu_i$, where $U_{\alpha i}$
are the elements of the PMNS mixing matrix.  The probability of
finding a neutrino with flavor $\beta$ is given by $\langle
\nu_\beta | \rho |\nu_\beta \rangle$. Hence, the oscillation
probability $\nu_\alpha \rightarrow \nu_\beta$ in vacuum is equal
to
\be
\label{eq:prob}
P_{\alpha \beta} = \langle \nu_\beta |
\rho^{(\alpha)}(t) |\nu_\beta\rangle =  \sum_{ij} U_{\beta i}^*
U_{\beta j} \, \rho_{ij}^{(\alpha)}(t) \,,
\ee
where $\rho_{ij}^{(\alpha)}(t)$ is given by Eq.~(\ref{evolved}) with
$\rho_{ij}(0) = \rho^{(\alpha)}_{ij}(0) = U_{\alpha i} U_{\alpha
j}^*$, which corresponds to the initial state $\nu_\alpha$.

In this paper, we consider the most economic scenario that
describes all the data. As we will show in the following, only one
matrix $D_n$ with
\be
\label{pattern}
d_1=d_2\ne d_3
\ee
is sufficient. Note that a similar pattern exists between neutrino
masses ($m_1 \simeq m_2 \ne m_3$), so it will be inspiring to build a
model that links the patterns of $H$ and $D$. Eq.~(\ref{pattern})
leads to
\be \label{gamma}
\gamma_{12}=0 \ \ {\rm and} \ \
\gamma\equiv\gamma_{13}=\gamma_{32}\ .
\ee
%
In the ranges of energy and baseline ($L$) for which
$\Delta_{21}L = \Delta m_{21}^2 L/(2 E_\nu) \ll 1$ , the
oscillations due to $\Delta m^2_{21}$ can be neglected, and
Eq.~(\ref{eq:prob}) yields
\begin{eqnarray}
P_{\mu e}(\gamma,L) &=& P_{e \mu}(\gamma,L)= 2|U_{\mu 3}|^2 |U_{e
3}|^2\left[1- e^{-\gamma L} \cos (\Delta_{31}L) \right]\,,
\nonumber\\
P_{ee}(\gamma,L) &=& 1-2|U_{e3}|^2(1-|U_{e3}|^2)\left[1- e^{-\gamma L}
\cos (\Delta_{31}L) \right]\,,
\label{probabilities} \\
P_{\mu \mu}( \gamma,L) &=& 1-2|U_{\mu 3}|^2(1-|U_{\mu 3}|^2)\left[1-
e^{-\gamma L} \cos (\Delta_{31}L) \right] \,. \nonumber
\end{eqnarray}

Notice that although the new term $\mathcal{D}[\rho]$ explicitly
breaks the time reversal symmetry, still within the framework of
the two-neutrino oscillation the equality $P_{e\mu}=P_{\mu e}$
holds. In principle, the decoherence effects can give rise to the
CPT violation; however, in this paper we assume that the
decoherence effects in the neutrino and antineutrino sectors are
the same.

The energy dependence of the decoherence parameter $\gamma$ in
Eq.~(\ref{gamma}) is not known; it should follow from a
microscopic theory of decoherence. In the absence of such a theory we
assume a power law: $\gamma \propto E_\nu^{-r}$, and
for convenience parameterize it as
\be
\gamma = \frac{\mu^2}{E_\nu}\left( \frac{40~{\rm
MeV}}{E_\nu}\right)^{r-1}\,,
\label{gamma-bamma}
\ee
where 40~MeV is  the typical neutrino energy in LSND.
We will estimate the allowed ranges of
parameters $r$ and $\mu^2$ in sec.~\ref{sec:other}, and in particular,
show that all the data can be described if
\be\label{eq:r}
r=4 \,,
\ee
which we will use as the reference value in our estimations.  The fast
decrease of $\gamma$, and consequently, the decoherence effect with
energy, is the key feature of the proposed scenario which allows us to
reconcile the LSND result with results of other experiments. Notice
that for any value of $r\ne 1$, the Lorentz symmetry is explicitly
violated.

\subsection{Explaining LSND events}

Let us now discuss the interpretation of the positive LSND result
through soft decoherence. For the LSND parameters, $E \sim 40$ MeV
and $L \simeq 30$ m, the oscillation phase $\Delta_{31}L \sim 5
\cdot 10^{-3}$, so that the oscillation effect is negligible and
according to (\ref{probabilities}) the appearance probability is
equal to
\be
P_{\mu e}(\gamma,L) = 2|U_{\mu 3}|^2 |U_{e 3}|^2
\left(1- e^{-\gamma L} \right) \approx
|U_{e 3}|^2\left(1- e^{-\gamma L}\right).
\ee
Hence, the LSND signal is determined by the 1-3 mixing and the
degree of decoherence given by the factor $(1- e^{-\gamma L})$. In
the case of strong decoherence, $\gamma L \gg 1$, this factor
converges to its maximum and
\be
P_{\mu e}(\gamma,L) \approx |U_{e 3}|^2 \,.
\ee
The LSND signal is simply given by the 1-3 mixing parameter, and
therefore the LSND probability provides a lower bound on the 1-3
mixing:
\be |U_{e 3}|^2 \geq P_{\mu e}^\mathrm{LSND} =
(2.6 \pm 0.8) \cdot 10^{-3} \,.
\label{min13}
\ee
For a small decoherence effect, $\gamma L \ll 1$, we obtain
\be
P_{\mu e}(\gamma,L) \approx |U_{e 3}|^2 \, \gamma L \,.
\ee
Therefore $|U_{e 3}|^2$ can be as large as the upper bound obtained
from the reactor experiments and other observations, $(|U_{e
3}|^2)^\mathrm{upper}$.  In this case the LSND result gives a lower
bound on $\gamma$ at typical LSND energies:
\be
\gamma \geq  \frac{P_{\mu e}^\mathrm{LSND}}
{(|U_{e 3}|^2)^\mathrm{upper} L_\mathrm{LSND}}\,.
\ee

\begin{figure}[t] \centering
    \includegraphics[width=0.6\textwidth]{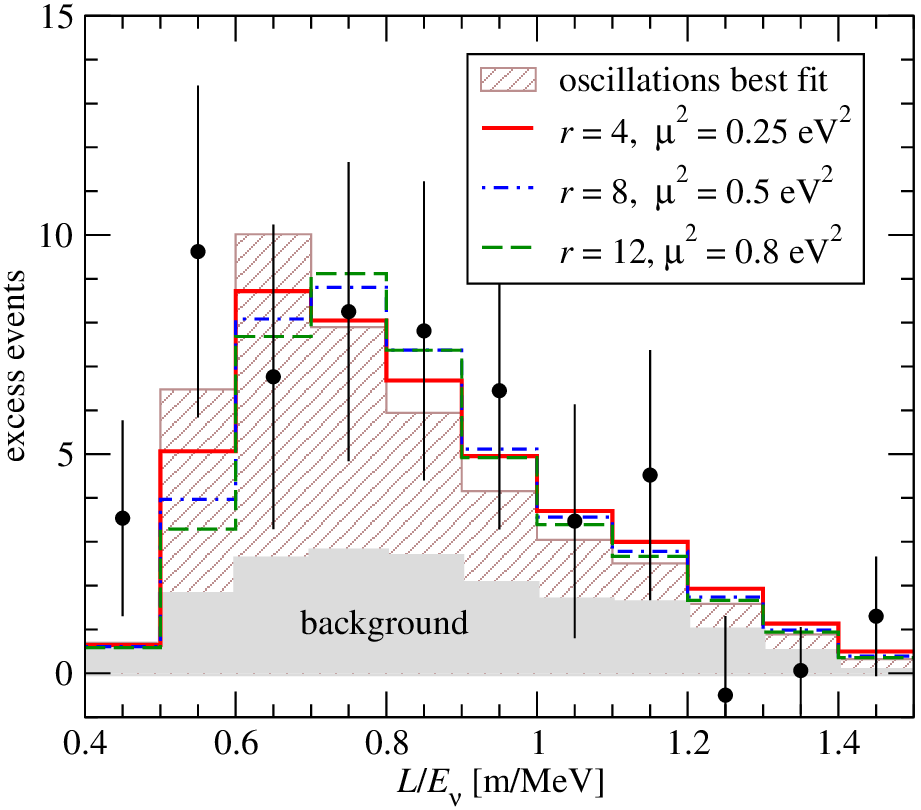}
    \mycaption{\label{spectrum} Spectrum of $\bar\nu_e$ excess events in
    LSND.  The histograms correspond to the prediction of the
    decoherence scenario for various values of $r$, see
    Eq.~(\ref{gamma-bamma}). The best fit value for $\mu^2$ in each case
    is assumed.}
\end{figure}

In Fig.~\ref{spectrum} we show the $L/E$ dependence of the excess of
the LSND events. The points with error bars are the LSND data and the
histograms show the prediction of the decoherence scenario (added to
the background) for various values of the power $r$, see
Eq.~(\ref{gamma-bamma}). To draw each histogram, the corresponding
best fit value for $\mu^2$ is assumed. For comparison we show the
spectrum expected due to oscillations in the model with a light sterile
neutrino. The decoherence reproduces the observed spectrum quite
well.  It leads to a softer energy spectrum than the oscillation
spectrum.  Furthermore, with increase of $r$ the softness increases
and the maximum shifts to lower energies.  Unfortunately these differences
are small and with the LSND statistics and uncertainties it is not
possible to distinguish decoherence and oscillation effects or
substantially restrict $r$ for $r\geq 4$.
For $\gamma L\sim 1$, $L \simeq 30$~m and typical LSND neutrino
energies $E_\nu \sim 40$~MeV, the relevant scale for the parameter
$\mu$ introduced in Eq.~(\ref{gamma-bamma}) is  $\mu^2\sim 0.1~
{\rm eV}^2$.

Thus, in the proposed scenario the LSND $\bar\nu_\mu\to\bar\nu_e$
signal is explained by the decoherence of the mass state $\nu_3$,
whose mixing with $\nu_\mu$ and $\nu_e$ are given by
$\cos\theta_{13}\sin\theta_{23}$ and $\sin\theta_{13}$,
respectively. The oscillation effect is negligible and plays no
role here.  Linking the LSND signal with the last unknown mixing
angle, $\theta_{13}$, is an exciting feature of the model.  We
will discuss the implications of this relation in
sec.~\ref{sec:future}.

\section{Reconciling LSND with other observations}
\label{sec:other}

To simultaneously accommodate the positive LSND signal and the lack of
any evidence for flavor transitions from other short-baseline
experiments, the value and energy-dependence of $\gamma$ have to be
properly chosen. In what follows, we demonstrate that since the
neutrino energy and baseline for each experiment are different, such a
choice is possible without affecting the successful description of
solar, atmospheric, KamLAND, and MINOS data in terms of neutrino
oscillations.

\subsection{Short baseline and reactor experiments}
\label{short-baseline}

\begin{figure}[t] \centering
    \includegraphics[width=0.6\textwidth]{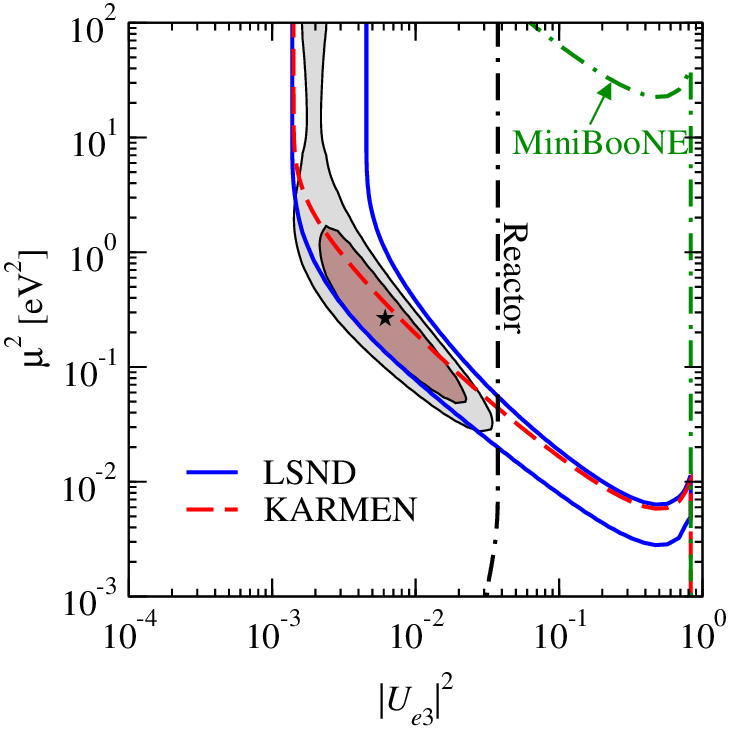}
    \mycaption{\label{fig:contours} Allowed regions in the plane of
    $|U_{e3}|^2$ and the decoherence parameter $\mu^2$ for $r = 4$.
    Shaded regions: global data at 90\% and 99\%~CL. Curves: 99\%~CL
    regions from LSND, KARMEN, MiniBooNE and reactor data (Bugey,
    Chooz, Palo Verde). The star marks the global best fit point. We
    marginalize over $\sin^2\theta_{23}$ taking into account the
    constraint from atmospheric neutrinos.}
\end{figure}

In short-baseline experiments, $\Delta m^2_{ij} L / (2E_\nu) \ll 1$
and we can neglect oscillations. Therefore, for a given power $r$ there are
three parameters to describe these experiments: $\mu^2$,
$|U_{e3}|^2 \equiv \sin^2\theta_{13}$ and $|U_{\mu3}|^2 \equiv
\cos^2\theta_{13}\sin^2\theta_{23}$ [see
Eq.~(\ref{probabilities})]. Hence, this scenario
involves the same number of parameters as the (3+1) sterile neutrino
oscillation schemes. We have performed a fit to the data from
LSND~\cite{Aguilar:2001ty} (decay-at-rest data), KARMEN~\cite{karmen},
MiniBooNE~\cite{MB, MB-data}, NOMAD~\cite{NOMAD},
CDHS~\cite{Dydak:1983zq}, and the reactor experiments
Bugey~\cite{Declais:1994su}, Chooz~\cite{Apollonio:2002gd}, and Palo
Verde~\cite{Boehm:2001ik}.  For technical details, see
Refs.~\cite{Maltoni:2007zf, Grimus:2001mn}. Results of the fit are
shown in the $\mu^2$-$|U_{e 3}|^2$ plane in Fig.~\ref{fig:contours}.

The LSND signal can be reconciled with the null-result from KARMEN
(which has the same neutrino energy as LSND) due to the somewhat
shorter baseline in KARMEN ($L \simeq 18$~m) and the exponential
dependence of the decoherence effect on the distance. The solid
and dashed curves in Fig.~\ref{fig:contours} show the allowed
regions from LSND and KARMEN, respectively. At 99\% C.L., there is
a significant overlap left, though there remains some tension
between the two results.  Notice that a similar tension exists in
other scenarios developed to explain the LSND results such as the
case of oscillations~\cite{0805.1764, Church:2002tc} or sterile
neutrino decay~\cite{Palomares-Ruiz:2005vf}, the latter having the
same exponential $L$ dependence as in the decoherence model.

The baseline and energy of neutrinos at MiniBooNE are one order of
magnitude larger than in LSND. As a result, the oscillation phase
$\Delta m^2 L/(2E_\nu)$ for the two experiments are of the same
order and oscillations can be neglected.  We find that for $r>2$,
at MiniBooNE $\gamma L \ll 1$ and therefore the  decoherence
effects are negligible, rendering the new effects at MiniBooNE
unobservable and explaining the non-observation of an appearance
signal.  As visible in Fig.~\ref{fig:contours}, for MiniBooNE the
decoherence becomes important only for $\mu^2 \gtrsim 20$~eV$^2$,
and is completely negligible for values of $\mu$ relevant for
LSND. Similarly to the case of standard oscillations, the soft
decoherence cannot account for the event excess observed in
MiniBooNE below 475~MeV. Therefore, following Ref.~\cite{MB}, we
rely on a yet to be identified explanation of this excess and use
only the data above 475~MeV in the analysis.

At the reference value $r = 4$ any decoherence effect for short
baseline experiments with energies substantially larger than
40~MeV is suppressed.  This implies no flavor transitions in the
``high energy'' experiments ($E_\nu \gtrsim 0.5$~GeV) MiniBooNE,
NOMAD, NuTeV, and CDHS.\footnote{A similar strategy to reconcile
low and high energy short-baseline data has been pointed out in
Ref.~\cite{Schwetz:2007cd} in the context of sterile neutrinos
with a non-standard energy dependence.} Therefore, our scenario is
trivially consistent with the null-result of these experiments.

The energies of reactor neutrinos are relatively low (of order 4~MeV). As
a result, for these neutrinos $\gamma$ is quite large:
\be\label{eq:gamma-MeV}
\gamma\sim 2.5 \, \frac{\mu^2}{0.2~{\rm eV}^2} \,
\left( \frac{4~{\rm MeV}}{E_\nu}\right)^4~
{\rm cm}^{-1} \,.
\ee
Thus, already after a few centimeters coherence is completely
lost and the survival probability in Eq.~(\ref{probabilities})
becomes independent of the energy and baseline:
\be
P_{\bar{e}\bar{e}}^\mathrm{reactor} \simeq 1-2|U_{e3}|^2(1-|U_{e3}|^2) \,.
\label{reactor}\ee
Thus, the Chooz bound on $P_{\bar{e}\bar{e}}$, which is
derived by comparing the measured $\bar{\nu}_e$ flux at a distance of
1~km from the source with the estimated flux from the consideration of
the power of the reactor can be directly translated into an upper
bound on $|U_{e3}|^2$. The upper bound on $|U_{e3}|^2$ from the
combination of Chooz, Bugey and Palo Verde data is shown in
Fig.~\ref{fig:contours}. We find $|U_{e3}|^2 < 0.04$ at $3\sigma$
(1~d.o.f.).

The shaded regions in Fig.~\ref{fig:contours} show the results of the
global analysis of the short-baseline experiments. We include
$\sin^2\theta_{23}$ as a free parameter in the fit, taking into
account the standard constraint from the Super-Kamiokande atmospheric
neutrino data~\cite{sk-atm}, which is hardly affected by decoherence
(see next section for a more detailed discussion).  For $r = 4$ we find
the best fit point at
\be\label{eq:bfp}
|U_{e3}|^2 = 6.1\times 10^{-3}  \,,\qquad
\sin^2\theta_{23} = 0.5         \,,\qquad
\mu^2 = 0.27\,{\rm eV}^2 \,,
\ee
with $\chi^2_\mathrm{min} = 89/(107-3)$~d.o.f. For these values of
the parameters, the averaged probability in LSND equals $P_{\mu
e}^\mathrm{LSND} = 2.3 \cdot 10^{-3}$ which is within one sigma of
the experimental value $(2.6 \pm 0.8) \cdot 10^{-3}$. The allowed
range for $|U_{e3}|^2$ is
\be
\label{eq:Ue3}
(1.4)\, 2.3 \times 10^{-3} < |U_{e3}|^2 < 2.1\,(3.4)\times 10^{-2}
\qquad \mbox{at 2 (3)$\sigma$ (1 d.o.f.)} \,.
\ee
Here the lower bound follows from the LSND result in the limit of
strong decoherence, see Eq.~(\ref{min13}). The upper bound comes from
the reactor experiments.

Let us evaluate the quality of the fit in more detail, and compare
it to the case of sterile neutrino oscillations in a (3+2)
scheme~\cite{Sorel:2003hf}. To this aim we divide the data into
sub-sets and check the consistency of these data sets by using the
so-called Parameter Goodness-of-fit (PG)
criterion~\cite{Maltoni:2002xd, Maltoni:2003cu}. It is based on
the $\chi^2$ function
\begin{equation} \label{eq:PG}
    \chi^2_\text{PG} =
    \chi^2_\text{tot,min} - \sum_i \chi^2_{i,\text{min}} \,,
\end{equation}
where $\chi^2_\text{tot,min}$ is the $\chi^2$ minimum of all data sets
combined and $\chi^2_{i,\text{min}}$ is the minimum of the data set
$i$. This $\chi^2$ function measures the ``price'' one has to pay by
the combination of the data sets compared to fitting them
independently. It should be evaluated for the number of d.o.f.\
corresponding to the number of parameters in common to the data sets,
see Ref.~\cite{Maltoni:2003cu} for a precise definition.

\begin{table}[t] \centering
    \begin{tabular}{l@{\qquad}cc@{\qquad}cc}
    \hline\hline
        & \multicolumn{2}{c@{\qquad}}{(3+2) oscillations}
    & \multicolumn{2}{c}{decoherence}
        \\
    Data sets & $\chi^2_\mathrm{PG}/$d.o.f. & PG
              & $\chi^2_\mathrm{PG}/$d.o.f. & PG
    \\
    \hline
        LSND vs NEV
    & 21.2/5 & 0.08\%
    & 8.6/2 & 1.4\%
    \\
    App vs Disapp
    & 17.2/4 & 0.18\%
    & 0.6/2 & 74\%
    \\
    \hline\hline
    \end{tabular}
    \mycaption{\label{tab:PG} Consistency of LSND versus all other
    no-evidence short-baseline data (NEV), and appearance versus
    disappearance short-baseline data for (3+2) oscillations and the
    decoherence model. We give $\chi^2_\mathrm{PG}$ according to
    Eq.~(\ref{eq:PG}) and the corresponding probability (``PG''). The
    results for (3+2) are taken from Tab.~3 of
    Ref.~\cite{Maltoni:2007zf}.}
\end{table}

First we test the consistency of LSND with all the other
null-result short-baseline experiments (NEV). The numbers given in
Tab.~\ref{tab:PG} show that for (3+2) oscillations LSND is
consistent with NEV only with a probability of 0.08\%, whereas in
the decoherence scenario the probability improves to 1.4\%. Note
that in the (3+2) case short-baseline data depend on 7 parameters,
whereas for decoherence only 3 parameters are available to fit the
data.\footnote{As before we work at fixed $r = 4$ and do not
consider the energy exponent as a free parameter.} The reason for
the still relatively low probability of 1.4\% is the
aforementioned tension between the LSND and KARMEN results.  In
order to illustrate this effect we perform a second test, dividing
the data into appearance experiments (LSND, KARMEN, NOMAD,
MiniBooNE) and disappearance experiments (CDHS, Bugey, Chooz, Palo
Verde).  In this approach, LSND and KARMEN data are summed into
the same data set and therefore, by assumption they are taken to
be consistent. As a result, the remaining tension between them
does not show up in the PG test.  Note that it is reasonable to
combine LSND and MiniBooNE into the same data set, because in both
cases considered here they are consistent: for (3+2) oscillations
they can be reconciled~\cite{Maltoni:2007zf} by invoking CP
violation~\cite{Palomares-Ruiz:2005vf, Karagiorgi:2006jf}, whereas
in the decoherence scenario the energy dependence of $\gamma$
guarantees the null-result of MiniBooNE. As shown in
Tab.~\ref{tab:PG}, we find an excellent fit in the decoherence
model (PG of 74\%), whereas (3+2) oscillations suffer from a sever
tension between appearance and disappearance experiments, allowing
for compatibility with a probability of only 0.18\%.

In summary, the soft decoherence proposed here provides an
excellent fit to short-baseline experiments, allowing for full
consistency of LSND and MiniBooNE, as well as appearance and
disappearance experiments. Only the well-known tension between LSND
and KARMEN remains unresolved.


\subsection{Other phenomenological consequences}
\label{sol+atm}

In this section, we show that our scenario is compatible with the
standard description of the solar~\cite{sno}, KamLAND~\cite{kamland},
atmospheric~\cite{sk-atm}, K2K~\cite{K2K} and MINOS~\cite{MINOS}
data in terms of neutrino oscillations.

\bigskip

\textbf{Solar and long-baseline reactor neutrino data:} In
Ref.~\cite{Fogli}, decoherence effects on solar and KamLAND neutrino
data have been studied (see also Ref.~\cite{Schwetz:2003se} for the
case of KamLAND only). The decoherence scenario  in \cite{Fogli}
differs from the one in the present paper: While we take $\gamma_{12} =0$
($d_1=d_2$) and discuss the effects of $\gamma=(d_1-d_3)^2$, the
authors of \cite{Fogli} focused on the effects of
$\gamma_{12} \equiv (d_1-d_2)^2$.
Our assumption $d_1=d_2$ in Eq.~(\ref{pattern}) ensures that
oscillations in the 1-2 sector are not affected by decoherence. This
leaves the dominant oscillations due to $\Delta m^2_{21}$ and
$\theta_{12}$ unchanged and guarantees the standard oscillation
explanations for the solar and KamLAND data.
Within our scenario, the effects of the damping factor on the solar
neutrino flux and KamLAND neutrinos are suppressed by $|U_{e3}|^2$.
The current uncertainties do not allow to resolve the effects of
$|U_{e3}|^2$. Moreover, in these experiments oscillations due to
$\Delta m^2_{31}$ are completely averaged out  and therefore
decoherence effects in the 1-3 sector are
unobservable.

We can use results of \cite{Fogli} to put an upper bound on
$\gamma_{12} = (d_1-d_2)^2$.  Writing $\gamma_{12} \equiv \gamma_0
(1\,{\rm GeV}/E_\nu)^r$, in Ref.~\cite{Fogli} bounds on $\gamma_0$
have been derived from solar and KamLAND data assuming $r=0,\pm 1,
\pm 2$. Extrapolating these results to $r=4$, one finds $\gamma_0
< 10^{-32}$~GeV. Using the parametrization shown in
Eq.~(\ref{gamma-bamma}), one has $\gamma_0 = 6.4\cdot 10^{-23}
\,{\rm GeV}\, (\mu_{12}^2/{\rm eV}^2)$, and
\be
\mu_{12}^2<10^{-10}~{\rm eV}^2\ll \mu^2 \,.
\ee

\bigskip

\textbf{Atmospheric neutrinos:} First, we note that because of the
smallness of $|U_{e 3}|^2$, the decoherence effects do not
considerably change the $\nu_e$ flux at low energies where the
Earth matter effect can be neglected. For high energies (multi-GeV
sample) where the 1-3 mixing is enhanced the decoherence effect
becomes negligible.  However, the muon neutrino disappearance
probability can be significantly affected. In order to have a
detectable effect of decoherence, the neutrino energies and
baselines should be in a range for which $\gamma L \gtrsim 1$ and
$\Delta_{31}L \lesssim 2\pi$. For $\Delta_{31}L \gg 2 \pi$,
regardless of the value of $\gamma$, the interference term
$e^{-\gamma L}\cos (\Delta_{31}L)$ averages out and the
sensitivity to the new effects is lost.  Putting the two
conditions together, we find that the effects can be noticeable
only for $E_\nu< 400$~MeV. On the other hand, for $E_\nu<200$~MeV,
the produced muons cannot be detected in Super-Kamiokande.

\begin{figure}[t] \centering
    \includegraphics[width=0.9\textwidth]{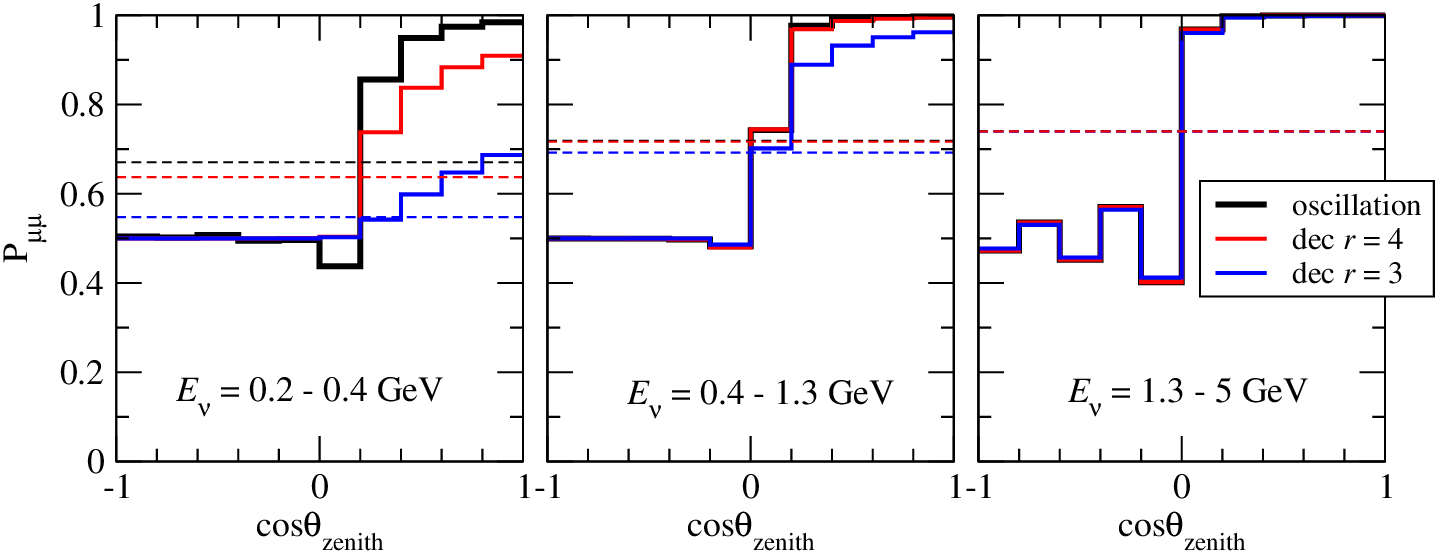}
    \mycaption{\label{atm} The zenith angle dependence of the
    $\nu_\mu$ survival probability relevant for atmospheric neutrinos
    in three different energy intervals for $r = 3, 4$ and for the
    case of oscillations without decoherence. The horizontal dashed
    lines show the corresponding probabilities averaged over all
    zenith angles.}
\end{figure}

In Fig.~\ref{atm} we show dependence of the survival probability
$\nu_\mu \rightarrow \nu_\mu$ on the neutrino zenith angle for
different energy ranges. In the lowest energy range, $200~{\rm
MeV}< E_\nu <400~{\rm MeV}$, the oscillation length
$L_\mathrm{osc} = 2\pi E_\nu / \Delta m^2_{31} \sim 100$~km.
Hence, the condition $\Delta_{31} L \sim \pi$ implies that only
for neutrinos arriving from above (for zenith angles smaller than
$90^\circ$) the damping effects are significant.  As follows from
Fig.~\ref{atm}, in the vertical direction for $r = 4$ and $r = 3$
the effect can reach 10\% and 30\%, correspondingly.  However, for
these low energies the direction of the initial neutrino is not
related to the muon direction and hence, the distribution of
$\mu$-like events is averaged over the zenith angle. As a result,
the decoherence leads to a zenith angle independent decrease of
the sub-GeV $\mu$-like events. The size of the effect is
illustrated by the horizontal dashed lines in Fig.~\ref{atm},
which correspond to the zenith angle averaged survival
probability. For $200~{\rm MeV}< E_\nu <400~{\rm MeV}$, we find a
suppression of 5\% (18\%) for $r=4$ ($r=3$).  This effect can be
detected as a decrease of the ratio of the $\mu$-like to $e$-like
events. We find that for $r \ge 4$ this decrease is below the 5\%
experimental uncertainty on this ratio~\cite{sk-atm}. Essentially
this effect determines the lower bound on $r$.  Let us note that
at these low energies, there is some excess of $e$-like events in
Super-K, which can be interpreted as renormalization and deficit
of $\mu$-like events and the latter can be explained by the
decoherence in our scenario.

In the energy interval $400~{\rm MeV}< E_\nu <1.3~{\rm GeV}$ the event
suppression due to the decoherence effect is below 5\% even for $r =
3$. The decoherence leads to a flattening of the zenith angle
dependence. However, the averaging over the zenith angle in the
$\mu$-like events---though incomplete---is still strong.  So
uncertainties in the extraction of the zenith angle in the
Super-Kamiokande experiment do not allow us to identify the effect of
decoherence for $r > 3$. In the multi-GeV range the decoherence effect
is strongly suppressed. In the Soudan experiment \cite{soudan}, in
addition to the energy of the produced muon, the recoil energy of
proton (in quasi-elastic interaction) is also measured. As a result,
the zenith angle of the incoming neutrino can be deduced. However, in
this experiment, the statistical error is larger than 10\% and the
decoherence effects cannot be resolved. We conclude that for $r \ge 4$
our scenario is consistent with the oscillation interpretation of
atmospheric neutrino data.\footnote{We thank Michele Maltoni for
communication on this point.}

\bigskip

\textbf{Long-baseline accelerator experiments:} At
MINOS~\cite{MINOS} experiment with $E_\nu \gtrsim 2$~GeV and
$L=730$~km, the damping effect is very small, $(1-e^{-\gamma L})
\approx \gamma L  <0.03$, and the present uncertainties do not
allow to resolve the effect. In the K2K experiment~\cite{K2K} the
energies of the neutrinos are smaller and the decoherence factor
is slightly larger. However, still the statistics below GeV is too
low for the experiment to be sensitive to the soft decoherence.

In Ref.~\cite{Lisi}, the decoherence effects in the atmospheric and K2K
neutrino data have been studied. However, the bounds derived in
\cite{Lisi} do not apply here because we assume a steeper decrease of
$\gamma$ with energy.

\bigskip

{\bf Radioactive source experiments:} In calibrations of the
Gallium solar neutrino detectors SAGE \cite{Abdurashitov:2005tb}
and GALLEX/GNO \cite{gallex} with artificial radioactive $^{51}$Cr
and $^{37}$Ar sources  deficits of the signals have been reported
in Ref.~\cite{Abdurashitov:2005tb}: the weighted average of the
ratio of the observed to expected event numbers equals
$R_\mathrm{Ga} = 0.88\pm 0.05$ \cite{Abdurashitov:2005tb}. This
result was interpreted as an indication of electron neutrino
disappearance due to oscillations into sterile
neutrinos~\cite{Giunti:2006bj}. For the energies of the
radioactive sources employed in these experiments, $E_\nu \sim
0.8$~MeV, the damping parameter $\gamma$ is huge [see
Eq.~(\ref{eq:gamma-MeV})] and the decoherence length $\sim
1/\gamma \sim 5 \cdot 10^{-4}$~cm. So, complete decoherence occurs
over a small fraction of a millimeter and the survival probability
is given by Eq.~(\ref{reactor}). For the best fit value of
$|U_{e3}|^2$ shown in  Eq.~(\ref{eq:bfp}), we find a small effect:
$P_{ee} \simeq 0.99$. However, if $|U_{e3}|^2$ is at its $3\sigma$
upper bound given in Eq.~(\ref{eq:Ue3}) we obtain $P_{ee} \simeq
0.93$. The latter is within the $1\sigma$ range of $R_\mathrm{Ga}$
and therefore the results of the calibration experiments can be at
least partially explained.

\section{Future tests of the scenario}
\label{sec:future}

\subsection{Reactors versus accelerators }

The possibility to test decoherence effects at future
long-baseline experiments has been discussed in~\cite{lbl, Sakharov}.
In our scenario, an explanation of the LSND signal implies a lower
bound on the mixing angle $\theta_{13}$.  Therefore, upcoming
oscillation experiments aiming at the measurement of this mixing angle
will provide a crucial test.

The next generation of reactor experiments like
Double-Chooz~\cite{double-chooz}, Daya Bay~\cite{daya} and
Reno~\cite{reno} will search for $\theta_{13}$ by comparing the
anti-neutrino flux measured at a near and far detector.  As discussed
in sec.~\ref{short-baseline}, a remarkable consequence of our
decoherence scenario is that for baselines larger than a few
centimeters the interference disappears, and $P_{\bar{e}\bar{e}}$ does
not vary with $L$ or $E_\nu$, see Eq.~(\ref{reactor}).  As a result,
the comparison of signals in near and far detectors at reactors will
not reveal oscillations. The model predicts a $\bar{\nu}_e$
flux reduction already at the near detector with respect to the
initial flux emitted from the reactor.  However, establishing this
reduction would rely on the ability to determine the original flux
with better than 1\% accuracy. This seems difficult to achieve.

In contrast to the reactors, in the future long-baseline accelerator
experiments T2K~\cite{T2K} and NO$\nu$A~\cite{NOVA} the damping
effect will be quite small, $(1 - e^{-\gamma L}) \simeq \gamma L \sim
0.02$, due to smallness of $\gamma$ in the GeV energy
range. Consequently, the sensitivity of these experiments to $|U_{e3}|^2$
through measurements of $P_{\mu e}$ will be basically unaffected by
decoherence. The sensitivity of T2K and NO$\nu$A is in the range
$\sin^22\theta_{13} \gtrsim 0.01$, and therefore these experiments can
probe the best fit point of our scenario, Eq.~(\ref{eq:bfp}), and
large part of the allowed range Eq.~(\ref{eq:Ue3}), though the lower
end of the 3$\sigma$ interval for $|U_{e3}|^2$ may escape detection.
Hence, by comparing the results of T2K and the upcoming reactor
experiments, the validity of the present scenario can be tested. The
following three situations are of particular interest.
\begin{itemize}
\item
Comparing the flux at far and near detectors, the reactor
experiments such as Double-Chooz and Daya Bay would establish a
non-zero value of $|U_{e3}|^2$ in agreement with the value
extracted from T2K and NO$\nu$A measurements of $P_{\mu e}$. In
this case, our scenario would be ruled out because, as discussed
above, $P_{\bar{e}\bar{e}}$ at the near and far detectors should
be the same.

\item
Neither the reactor experiments nor the long baseline experiments
find any evidence for nonzero $|U_{e3}|^2$ and put an upper bound
of 0.0025 on its value ($\sin^22\theta_{13} < 0.01$). In this
case, the constraints on our scenario become quite tight, and the
allowed region for $|U_{e3}|^2$  shifts to the lower end of the
$3\sigma$ interval given in Eq.~(\ref{eq:Ue3}). This shifts
$\mu^2$ to larger values, of order 1~eV$^2$ (see
Fig.~\ref{fig:contours}) and increases the tension between LSND
and KARMEN data. In order to fully rule out the model, the
3$\sigma$ bound on $|U_{e3}|^2$ has to be pushed below 0.0014
($\sin^22\theta_{13} < 0.0056$), which probably requires to go
beyond the initial phases of T2K and NO$\nu$A.

\item
While comparing the reactor neutrino fluxes at near and far detectors
reveals no evidence for missing $\bar{\nu}_e$, T2K and/or NO$\nu$A
consistently report a value of $|U_{e3}|^2$ in the range
$(0.0014,0.034)$. This situation cannot happen within the standard
oscillation scenario. Thus, such an outcome can be considered as a
strong hint in favor of our scenario.
A way to confirm or refute this hint is to compare the measured
reactor neutrino flux with the original flux estimated from the
power considerations, as it had been done in the analysis of the
CHOOZ data.  Of course, to do this the systematical uncertainties
in the flux estimations have to be overcome.
\end{itemize}

Currently the MiniBooNE experiment is taking data in the
anti-neutrino mode. Since our model invokes neither CP nor CPT
violation, the prediction for anti-neutrinos is the same as for
neutrinos. Hence one expects a null-result for MiniBooNE
anti-neutrino run. For the low energy experiment proposed in
Ref.~\cite{Grieb:2006mp} using the LENS detector, the situation is
similar to the one in the Gallium calibration experiments mentioned
above. After very short distances decoherence sets in, leading to
a constant event suppression according to Eq.~(\ref{reactor}). No
distance dependent effect would be observed, and the measurement
has to rely on the comparison of expected and predicted numbers of
events, which might be difficult due to the normalization
uncertainties.

\bigskip

The phase I of T2K is planned to be followed by a phase II which
can probe the effects of $\Delta m_{21}^2$ and, for relatively
large values of $|U_{e3}|$ as predicted in our model, measure the
Dirac CP-violating phase~\cite{T2K}. The energy of the neutrino
flux in the second phase of T2K will be around 750 MeV for which
$\gamma \sim \Delta m_{21}^2/E_\nu$. Thus,  the decoherence
parameter is $(1 - e^{-\gamma L}) \simeq \gamma L \sim \Delta
m^2_{21}/\Delta m^2_{31} \simeq 0.03$. Hence, one expects a
distortion of the energy spectrum at the level of a few percent.
The effect might be difficult to observe in the appearance signal,
since in this case, the number of events expected in T2K-II is of
order of 1000, {\it i.e.}, the statistical error is  a few
percent. Moreover, from Eqs.~(\ref{probabilities}), we observe
that the decoherence effects on the appearance probability,
$P_{\mu e}$, is further suppressed with a factor of $|U_{e3}|^2$.
However, a spectral distortion due to decoherence is also expected
for the $\nu_\mu$ disappearance signal, where statistics is much
larger.

Future projects for superbeam (SPL) or Beta Beam experiments from
CERN to a megaton scale detector in Frejus~\cite{Campagne:2006yx}
have very good sensitivity to probe our scenario. These
experiments will have neutrino energies roughly a factor two
smaller than T2K, which enhances $\gamma$ by a factor $2^r$ and
therefore, $\gamma L \sim \mathcal{O}(1)$. Thus, the experiments
would operate in the regime that the decoherence effect is
significant and this would lead to very different spectral
signatures as compared to standard oscillations.

\bigskip

Ref.~\cite{Raghavan} suggests to use so-called M\"ossbauer neutrinos
to measure $\theta_{13}$. The energy of such neutrinos is low (for
example, for neutrinos from Tritium decay, $E_\nu=18.60$~keV). For
such neutrinos, our scenario predicts Eq.~(\ref{reactor}), while
within the standard oscillation scenario,
$P_{ee}=1-4|U_{e3}|^2(1-|U_{e3}|^2)\sin^2
{\Delta_{31}L}/{2}$~\cite{Akhmedov:2008jn}. Considering that the
energy spectrum of the M\"ossbauer neutrinos is monochromatic, it will
not be possible to discriminate between the two scenarios by studying
the energy dependence of $P_{ee}$. However, by measuring the flux at
several distances it will be possible to make a distinction. If the
soft decoherence is realized in nature and the measurement is done at
a distance $L$ with $\sin^2 {\Delta_{31}L}/{2}\ne 1/2$, neglecting the
decoherence effects will cause a disagreement between the results of
T2K and NO$\nu$A with this measurement.

\subsection{Decoherence in matter; Supernova neutrinos}

In matter, the total Hamiltonian includes the interaction term
described by the matrix  of potentials $V$:  $H \rightarrow H +
V$. In the neutrino mass basis $V$ is non-diagonal, and therefore
$[H, D_n] \neq 0$. The fact that the decoherence matrix does not
commute with $V$ and consequently with the total Hamiltonian leads
to a new effect: {\it statistical equilibration} of flavors and
masses.  This means that after sufficient time
($t\stackrel{>}{\sim}([V,D])^{-1/2}$), $\rho$ converges to unit
matrix times a normalization factor  and consequently, in the case
of two neutrino mixing, the probabilities of finding neutrinos
with the mass $m_1$ and $m_2$  in the course of evolution converge
to $P_1 = P_2 = 1/2$. Similarly  for mixed flavors as a result of
long enough evolution $P_e = P_x = 1/2$, where $x$ is some
combination of $\nu_\mu$ and $\nu_\tau$ neutrinos with which
$\nu_e$ mixes. This phenomenon is similar to the flavor
equilibration in the presence of mixing and inelastic collisions
which destroys coherence. Such a type of equilibration has been
considered for the active-sterile neutrino oscillations in the
Early Universe.

In our scenario non-trivial interplay of the decoherence and matter
effect should take place for solar, atmospheric and supernova
neutrinos. No significant effect is expected for the solar neutrinos
inside the Sun as well the Earth and also for the atmospheric
neutrinos inside the Earth. Indeed, for solar neutrinos due to low
energies the matter effect on 1-3 mixing is negligible.  For
atmospheric neutrinos inside the earth the matter effect on 1-3 mixing
is substantial in the multi GeV range where decoherence is negligible.
In contrast, for supernova neutrinos ($E = 5 - 40$ MeV, and huge
densities) both matter and decoherence effects are strong.

Let us estimate qualitatively the decoherence effect on supernova
neutrinos. In our scenario the 1-3 mixing should be relatively large
so that without decoherence the conversion in the H-resonance region
(due to 1-3 mixing and mass splitting) should be highly adiabatic. In
the case of normal mass hierarchy that would lead to the transition
$\nu_e \rightarrow \nu_3$ inside the star. As a consequence, no earth
matter effect is expected in the neutrino channel. In contrast, in the
presence of decoherence and equilibration of mass states only half of
$\nu_e$'s will end up as $\nu_3$ and another half will transform in
the H-resonance region to $\nu_{2m}$ - the second eigenstate in
matter.  Soft decoherence leads to features in supernova neutrinos
similar to the ones expected within standard oscillations with $\sin^2
2\theta_{13} < 10^{-4}$; {\it i.e.,} non-adiabatic conversion in the
H-resonance~\cite{Dighe}. In particular, the earth matter effect
should show up.  Thus, one expects mismatch of the 1-3 mixing measured
at the accelerators and bound from studies of supernova neutrinos.

Detailed analysis of the decoherence in matter and effects on
supernova neutrinos are beyond the scope of this paper. Here we
will present a simplified consideration which  shows that the
statistical equilibration is achieved already before the 1-3
resonance. Recall that in matter the mass states are not the
eigenstates of propagation and therefore oscillate. In particular,
$\nu_3$ should oscillate into a certain combination of $\nu_1$ and
$\nu_2$, $\nu_a$, which depends on the density. In our soft
decoherence scenario, the decoherence length for supernova
neutrinos ($E \sim (10 - 20)$ MeV) is rather small: $L_{\rm decoh}
= 1/\gamma \sim 1$ m. The oscillation length in matter is
determined by the refraction length: $l_m \approx 2\pi/V$.
Therefore for not very large $V$ we have $L_{\rm decoh}  \ll l_m$.
In this limit, the oscillation effect can be  considered in the
following way. The neutrino trajectory can be divided into
intervals of  size $L_{\rm decoh}$. A given mass state, $\nu_3$
oscillates on the first interval $L_{\rm decoh}$ $\nu_3
\rightarrow  \sqrt{1 - \alpha_1^2} \nu_3 + \alpha_1 \nu_a$, where
$|\alpha_1|^2 = P_1$ is the transition probability. At the end of
this interval the coherence between $\nu_3$ and $\nu_a$ components
of the state  is destroyed and in the next interval, $L_{\rm
decoh}$, they will oscillate independently. At the end of the
second interval we will have split of the states again, and so
forth. The probability of transition $\nu_3 \rightarrow \nu_a$ in
the $i$th interval,  $P_i \ll 1$ can be estimated as 
\be 
P_i \sim
\sin^2 2 \theta^m_{\rm mass}(V_i) \, \sin^2 \phi_i \approx 
\sin^2 2 \theta^m_{\rm mass}(V_i) \, \phi_i^2 \approx 
\sin^2 2\theta_{13} \, (V_i L_{\rm decoh})^2 \,, 
\ee 
where $\theta^m_{\rm mass} = \theta_{13}^m - \theta_{13}$ is the mixing
angle of the mass states in matter ($\theta_{13}^m$ is the mixing
angle of the flavor states in matter), $V_i$ is the matter potential
in the $i$th interval, $\phi_i$ is the half-phase of oscillations in
the $i$th interval .  Notice that to derive Eq.~(23) we have used the
constant density approximation for the oscillation probability within
each interval of size $L_{\rm decoh}$. We have then used $\phi_i \ll
1$ which follows from $L_\mathrm{decoh} \ll l_m$.

Since  the initial  flux is mainly composed of $\nu_3$, the flux
transition $\nu_3 \rightarrow \nu_a$ will dominate over the
opposite transition $\nu_a \rightarrow \nu_3$,  and eventually
this will lead to the equilibration of the $\nu_3$ and $\nu_a$
fluxes. The total transition probability after passing $n$
intervals is given by
\be 
P \approx \sum_i^n P_i = \sum_i^n \sin^2
2\theta_{13} V_i^2 L_{\rm decoh}^2 \,. \label{probbb} 
\ee 
This formula is valid when $P \ll 1$ (linear regime) for which the
inverse transition still can be neglected. Still the condition $P \sim
1$ allows to evaluate the length (numbers of intervals) over which the
equilibration is achieved. Substituting summation in (\ref{probbb}) by
integration we obtain 
\be P \approx \sin^2 2\theta_{13} L_{\rm decoh}
\int_{r_0}^{r_R} V^2(r) dr \,,
\label{probint} 
\ee 
where $r_R \sim 10^7$ m is the radius of the
resonance layer. (Recall we are estimating the distance from the
resonance layer in the direction of center of a star on which
equilibration is reached.) Taking $V = V_R (r_R/r)^3$ we obtain
from (\ref{probint}) 
\be P  \approx \sin^2 2\theta_{13} [L_{\rm
decoh} V(r_0)]^2 \frac{r_0}{5 L_{\rm decoh}} \,. \label{probest} 
\ee
Taking $V_R\sim {\Delta m_{13}^2}/{E_\nu}$, from Eq.~(\ref{probint}),
we find that for the whole range of $\theta_{13}$ within our scenario
[see Eq.~(\ref{eq:Ue3})], propagating from $r_0\sim 0.1r_R$ to
$r_R$ the equilibration condition ($P\sim 1$) is fulfilled.

\section{Conclusions}
\label{sec:conclusions} 

Clearly there is no simple explanation of the LSND result which is
consistent with other neutrino data. According to some proposals it
requires the combination of two exotic mechanisms. One can ask if any
yet unknown mechanism exists which can provide a description of
(reconcile) all the data. One can take the bottom up approach and try
to uncover properties (energy and distance dependence) of this unknown
mechanism. Our proposal is essentially along this line.

We have proposed an explanation of the LSND signal as manifestation of
the quantum decoherence of mass states associated to the 1-3 mixing
and mass splitting.  The decoherence leads to a complete or partial
damping of the interference terms in the oscillation
probabilities. Our phenomenological scenario is based on the
three-neutrino framework (without any sterile neutrinos) and makes use
of a single decoherence parameter $\gamma$ with a sharp energy
dependence (soft decoherence). The main features of our scenario,
which allow us to reconcile the LSND signal with the results of other
experiments are:
\begin{itemize}
\item zero or negligible decoherence effect on the 1-2 mixing and splitting
  ($\gamma_{12} = 0$);
\item decoherence of only the $\nu_3$ mass eigenstate;
\item a strong decrease of the decoherence effect with the neutrino
  energy:\\
  $\gamma \equiv \gamma_{13} = \gamma_{23} \propto E_\nu^{-r}$.
\end{itemize}
The strong decrease of $\gamma$ with energy allows us to accommodate
the LSND signal, while being consistent with the null-results of
experiments at higher energies, such as MiniBooNE, CDHS, NOMAD, and
NuTeV.  At the same time this energy dependence guarantees standard
neutrino oscillations in the atmospheric and MINOS
experiments. The lower bound on the energy exponent, $r \gtrsim 4$,
follows from the low energy atmospheric neutrino data; $r = 4$ can be
considered as an optimal reference value. The assumption $\gamma_{12}
=0$ ensures that oscillations in the 1-2 sector relevant for solar
neutrinos and the KamLAND experiment are not affected.

The LSND signal is determined by the mixing angle $\theta_{13}$,
and the degree of decoherence. It implies a lower bound on 1-3
mixing, $\sin^2 2\theta_{13} > 0.006$ ($3\sigma$), which
corresponds to complete decoherence.  This in turn, leads to
testable predictions for upcoming experimental searches of 1-3
mixing. For reactor experiments (MeV energies) like Double-Chooz
or Daya Bay we predict full decoherence already after a few
centimeters, leading to no oscillation effect when
 results from near and far detectors are compared.  In contrast,
the long-baseline accelerator experiments (GeV energy range) like
T2K or NO$\nu$A are practically not affected by decoherence and a
signal for $\theta_{13}$ should show up. Hence, a mismatch in the
$\theta_{13}$ measurements of upcoming reactor and accelerator
long baseline experiments would be a clear indication for the
proposed scenario. At low-energy long-baseline experiments such as
the CERN SPL superbeam or a Beta Beam with a relativistic
$\gamma$-factor of 100, our decoherence scenario will lead to a
distinct energy spectrum of the appearance as well as
disappearance signals. The soft decoherence can also show up as a
distortion in the energy dependence of disappearance probability
in the second phase T2K.
This scenario can also partially account for the anomaly found in
Gallium radioactive source experiments~\cite{Abdurashitov:2005tb},
though explaining the full effect might be difficult. Soft decoherence
can also affect the supernova neutrinos. Despite relatively large
$\theta_{13}$, decoherence leads to a neutrino composition similar to
the case of non-adiabatic conversion which takes place in the case of
the standard oscillation with $\sin^22\theta_{13}<10^{-4}$. Thus,
comparing $\theta_{13}$ measured by T2K and NO$\nu$A and supernova
bound on this mixing angle can be considered as another way to test
the present scenario.

The energy dependence of the decoherence effect has to be explained by
an underlying theory that gives rise to quantum decoherence. In this
paper we considered the simplest power law dependence of $\gamma$ in the
whole energy range. In general $\gamma$ may have a more complicated
dependence. Indeed, the fact that $\gamma \to \infty$ for $E_\nu \to 0$
indicates that there should be some low energy cut-off below which the
dependence of $\gamma$ on $E_\nu$ becomes modified.  The restrictions
on $\gamma$ come from neutrino data in the energy range from $\sim
10$~MeV to multi-GeV. With a general energy dependence, it is possible
that for $E_\nu<10$~MeV, $\gamma$ remains constant or even becomes
small again. Such a behavior could modify our predictions for low
energy experiments, especially reactor neutrinos.
Note, however, that under the power law assumption the coherence for
reactor neutrinos is lost already within a few cm, whereas the close
detectors are at several hundred meters.  So, if $\gamma$ is constant
below LSND energies or even decreases not very fast, we still will
have decoherence.  Hence, most probably at least partial decoherence
will be observed and in this case still 1-3 mixing will be different
in reactor and accelerator experiments.  Only in the case of a very
sharp cut-off below $E_\nu \sim 10$~MeV (a behavior which appears
quite unnatural) one will see the same 1-3 mixings in both cases.  In
any case the decoherence can be still probed by studies of the energy
spectrum in the phase II of T2K, and CERN beta beam and SPL
experiments. Moreover, we still expect a disagreement between
$\theta_{13}$ measurements by T2K and NO$\nu$A and the bounds from
supernova data.

\subsection*{Acknowledgments}

We thank Michele Maltoni for useful communications. A.Yu.S.\
acknowledges some early discussions of the decoherence and LSND result
with Srubabati Goswami. Y.F.\ thanks ICTP where part of this work has
been done for its support and the hospitality of its staff. She is
also grateful to Ashoke Sen for useful discussions. T.S.\ would like
to thank Alexander Sakharov for discussions on decoherence due to
quantum gravity.


\end{document}